\documentclass[prc,aps,showpacs,preprintnumbers,amsmath,amssymb]{revtex4}
\usepackage{epsfig}
\usepackage{graphicx}
\usepackage{sublabel}
\usepackage{verbatim}
\def\Vec#1{\mbox{\boldmath $#1$}}

 \newcommand\beq{\begin{equation}}
 
 \newcommand\eeq{\end{equation}}
 \newcommand\beqn{\begin{eqnarray}}
 \newcommand\eeqn{\end{eqnarray}}

\def\Vec#1{\mbox{\boldmath $#1$}}

\def\beq{\begin{equation}}
\def\eeq{\end{equation}}
\begin{document}

 \date{today}
\vskip 2mm
\date{\today}
\vskip 2mm
\title{
SEMI-INCLUSIVE DEEP INELASTIC SCATTERING OFF FEW-NUCLEON SYSTEMS:
TAGGING THE EMC EFFECT AND HADRONIZATION MECHANISMS WITH DETECTION
OF SLOW RECOILING NUCLEI}
\author{C. Ciofi degli Atti}
\author{L. P. Kaptari}
\altaffiliation{On leave from  Bogoliubov Lab.
      Theor. Phys.,141980, JINR,  Dubna, Russia; supported
through the program "Rientro dei Cervelli" of the Italian Ministry
of University and Research}
\affiliation{Department of Physics, University of Perugi,\\ and\\
   Istituto Nazionale di Fisica Nucleare, Sezione di Perugia,
   Via A. Pascoli, I-06123, Italy}

\begin{abstract}
\noindent The semi-inclusive deep inelastic scattering  of
electrons off  $^2H$ and $^3He$ with detection of slow protons and
deuterons, respectively, i.e. the processes $^2H(e,e'p)X$ and
 $^3He(e,e'd)X$, are calculated within the spectator mechanism, taking into account
 the final state interaction of the nucleon debris with the detected protons and
 deuterons. It is shown that by  a proper choice of the  kinematics the origin
  of the EMC effect and the details of the interaction between  the
 hadronizing quark and the nuclear medium can be investigated at a level
 which cannot be reached by inclusive deep inelastic scattering.
A comparison of the results of our calculations, containing no
adjustable parameters,
 with
recently available experimental data on the process $^2H(e,e'p)X$
shows a good agreement  in the backward hemisphere of the emitted
nucleons.
 Theoretical predictions at  energies that will be available at the upgraded Thomas Jefferson National Accelerator Facility
 are presented, and the possibility to investigate the
 proposed semi-inclusive processes at  electron-ion colliders is  briefly
 discussed.

\end{abstract}

\date{\today}


\maketitle

\maketitle
\large
\vskip 2cm
\newpage
\section{INTRODUCTION}
 In spite of many experimental and theoretical efforts
(for a recent   review see \cite{EMC}), the origin of the nuclear
EMC effect has not yet  been fully clarified and the problem as to
whether the quark distribution of nucleons undergoes  deformations
due to the nuclear medium remains open. Understanding
 the origin of the EMC effect  would be of great relevance in many
respects; consider,  for example, that
 most   QCD sum rules and predictions require
  the knowledge of the neutron quark distributions, which
 can only be extracted  from  nuclear experiments;
 this implies, from one side, a reliable  knowledge of  various non trivial nuclear
properties such as  the nucleon removal energy and momentum
distributions and,  from the other side,  a proper  treatment of
the lepton-nucleus reaction mechanism, including the effect of the
final state interaction (FSI) of the leptoproduced hadrons with
the nuclear medium.
 Since the dependence of the EMC effect upon the momentum transfer, $Q^2$,
 and the Bjorken scaling variable, $x_{Bj}$,
  is smooth,
 the measurements of the nuclear quark distributions
 in inclusive deep inelastic scattering (DIS) processes
have not yet determined enough constraints to
 distinguish between different theoretical approaches.
To progress in this field, one should go beyond inclusive
experiments, e.g.  by considering semi-inclusive deep inelastic
scattering (SIDIS) processes in which another particle  is
detected in coincidence with the scattered electron. The
"classical" SIDIS processes are the ones in which a fast hadron,
arising from the leading quark hadronization, is detected in
coincidence with the scattered electron. This type of SIDIS has
provided much information on hadronization in the medium (for a
recent review, see~\cite{brooks}), but not on the  EMC effect.
 An alternative type of SIDIS, namely the one in which, instead of the high energy leading hadron,
 a nucleus $(A-1)$ in the ground
state is detected in coincidence with the scattered
electron~\cite{scopetta}, has been shown to be very useful in
clarifying  the origin of the EMC effect and, at the same time, in
providing   valuable information on quark hadronization in the
medium, complementary to the information obtained, so far,  by the
analysis of the "classical" SIDIS process. In
Ref.~\cite{scopetta}, however,  the plane wave impulse
approximation (PWIA) was assumed to be the basic mechanism of the
process. A relevant step forward was made in Ref.~\cite{ciokop},
 where the theoretical approach was  extended by  considering   the
final state interaction of the hadronizing debris (the leading quark and the diquark)
 with the nucleons
of the nucleus $(A-1)$. This was done   within a theory of FSI
based on the eikonal approximation with the debris-nucleon
 interaction cross sections calculated by the hadronization model of Ref.~\cite{koppred}.
 In Ref.~\cite{ckk} this theory of FSI was applied to the treatment of
the process  $^2H(e,e'p)X$ in the limit of asymptotic values of
$Q^2$, whereas in Ref.~\cite{veronica} finite values of $Q^2$ were
considered. In the present paper the SIDIS processes $^2H(e,e'p)X$
and $^3He(e,e'd)X$ will be analyzed in details, presenting in the
former case a comparison with recent experimental
 data from Jlab~\cite{klimenko, kuhn}.  Since accurate nuclear wave functions,
 corresponding to realistic nucleon-nucleon (NN) interactions, e.g. the Urbana AV18
 interaction~\cite{AV18}, can be used for both the
 two- and three-body nuclei, our calculations can serve as a reference guide for calculations in complex nuclei.
 In  Sec. II the theoretical cross sections of  the  process $A(e,e'(A-1))X$, both in PWIA
 and with consideration of
 FSI, will be discussed; in Section~III, the SIDIS process  $^2H(e,e'p)X$  will
 be illustrated   and a comparison between theoretical calculations and experimental data will
  be presented; in Section IV the process ${^3}He(e,e'd)X$ is analyzed, illustrating
  how the SIDIS process we are considering could be used to tag the EMC effect and to investigate hadronization
  mechanisms.

\section{CROSS SECTION OF THE PROCESS  $A(e,e'(A-1))X$ WITH ACCOUNT OF FSI}

\begin{figure}[t]
\vskip 6cm
\begin{center}
\centerline{\hspace{0.5cm}
    \epsfysize=4.5cm\epsfbox{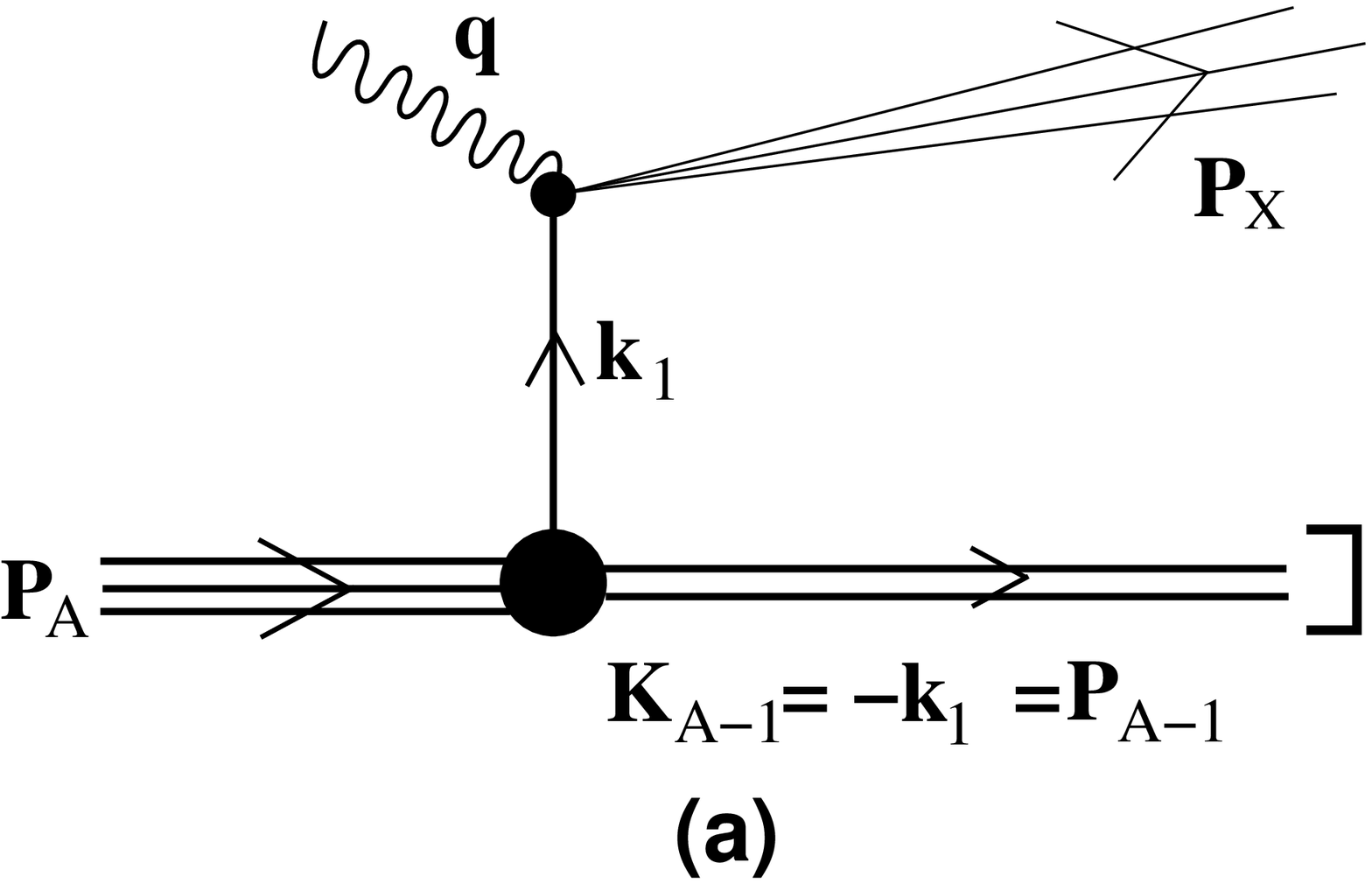}\hspace{0.5cm}
    \epsfysize=4.5cm\epsfbox{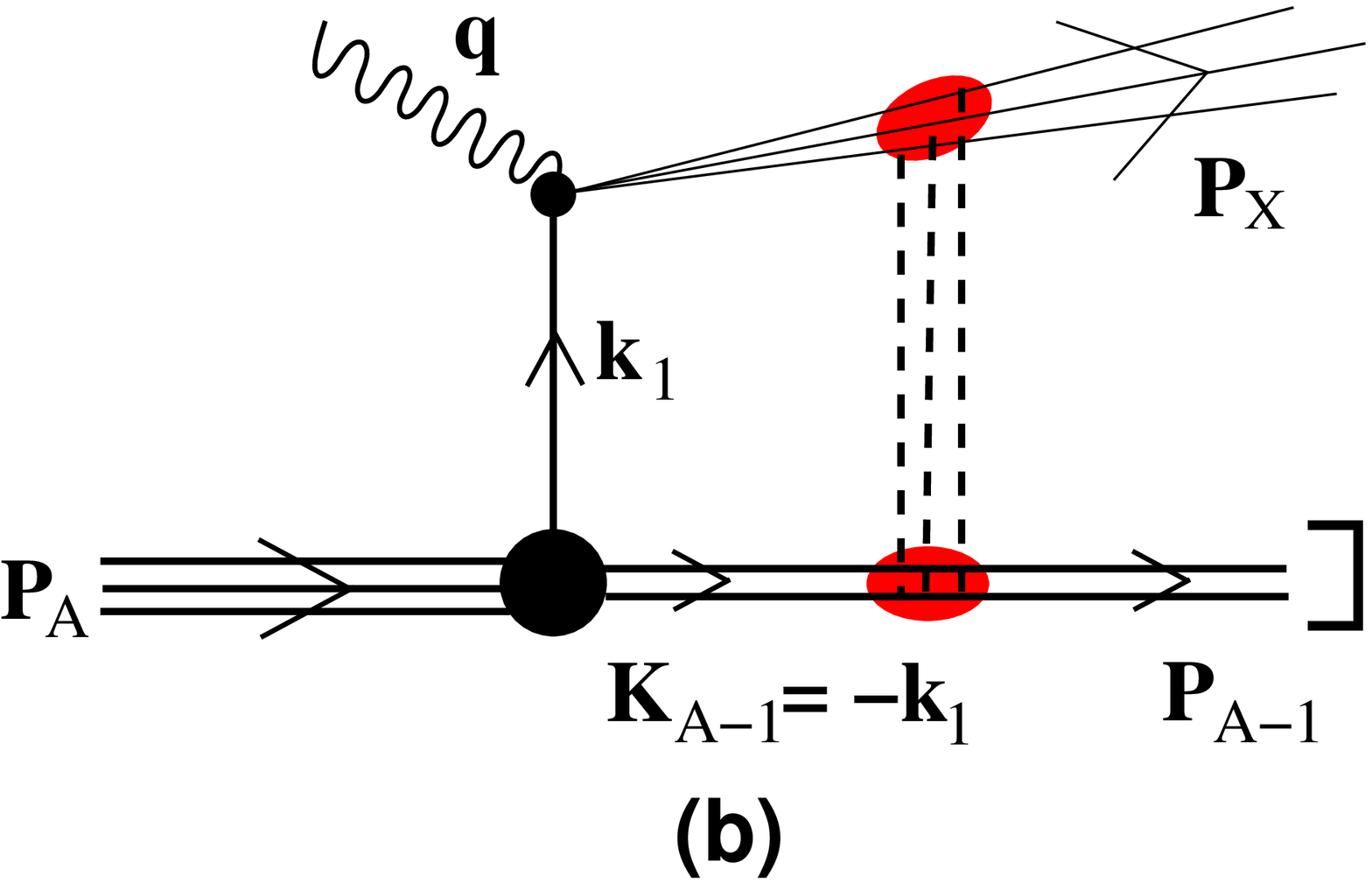}\hspace{0.1cm}}
    \caption{(Color online) The PWIA (a) and the FSI (b) contributions
    to the SIDIS process $A(e,e'(A-1))X$}
    \label{Fig1}
\end{center}
\end{figure}

The Feynman diagrams corresponding to the PWIA and FSI cross
sections are shown in Fig.~\ref{Fig1}. These diagrams describe the
so-called {\it spectator mechanism} in which  the virtual photon
hits a quark of a nucleon of the target $A$, and the nucleus
$(A-1)$ coherently recoils and is detected in coincidence with the
scattering electron. It is clear  that if the target nucleus is a
deuteron, the recoiling "nucleus", within the spectator mechanism,
can only be a nucleon, which, however, can in principle arise from
other mechanisms, e.g. current and/or target fragmentation. Target
fragmentation has been analyzed in Ref. \cite{veronica}, whose
conclusions will be briefly recalled in Section III; from now on,
we will only consider the spectator mechanism.
 If, on the contrary, the target is a nucleus with $A > 2$, the detection of
 an $(A-1)$ nucleus
 not only is strong evidence of the correctness of the spectator
mechanism, but also it can provide information on the nature of
the FSI between the nucleon  debris and $(A-1)$ nucleons.  In PWIA
the SIDIS differential cross section reads as
follows~\cite{scopetta, ciokop, veronica}:
\begin{eqnarray}
   &&\!\!\!\!\!\! \sigma^{A,PWIA} (x_{Bj},Q^2,|\Vec P_{A-1}|,y_A,z_1^{(A)})\equiv\sigma^{A,PWIA}=
  \frac{d\sigma^{A,PWIA}}{d x_{Bj} d Q^2  d \Vec   P_{A-1}}=\nonumber\\&&
   =
   K^A( x_{Bj},Q^2,y_A,z_1^{(A)})  z_1^{(A)}\,F_2^{N/A}(x_A,Q^2,k_1^2)\,n^A_0(|\Vec P_{A-1}|).
   \label{crosa-1}
   \end{eqnarray}
   Here $Q^2 =-q^2= -(k_e-k_e')^2 = \Vec q^{\,\,2} - \nu^2=4 {E}_e
{E}_e' sin^2 {\theta_e \over 2}$ is the four-momentum transfer
(with $\Vec q = \Vec k_e - \Vec k_{e'}$, $\nu= {E}_e - { E}_e' $
and $\theta_e\equiv \theta_{\widehat{\Vec k_e  \Vec k_{e'}}}$);
$y=\nu/{E}_e$;  $ x_{Bj} = Q^2/2m_N\nu $ is the Bjorken scaling
variable, with $m_N$ denoting the nucleon mass;
 $k_1 \equiv(k_{10},\Vec {k_1})$,
 with  $\Vec {k_1} \equiv - \Vec P_{A-1}
$, is the four momentum of the hit
 nucleon;
$F_2^{N/A}$ is the  DIS structure function of the nucleon $N$ in
the nucleus $A$, depending upon the nucleus scaling variable
$x_A$ and  $Q^2$ ({\it cf.} Eq.~(\ref{adef})); eventually,
  $K^A( x_{Bj},Q^2,y_A, z_1^{(A)}) $ is the following kinematical
  factor (note that in Ref. \cite{veronica} $y^A$, $z_1^{(A)}$, and $K^A$ were
  denoted $y_1$, $z_1$ and $K$, respectively)
   \begin{eqnarray}
   &&
   K^A( x_{Bj},Q^2,y_A,z_1^{(A)})=
   \frac{4\alpha_{em}^2}{Q^4} \frac {\pi}{x_{Bj}} \left( \frac{y}{y_A}\right)^2
   \left[\frac{y_{A}^2}{2} + (1-y_A) -
   \frac{k_1^2x_{Bj}^2 y_A^2}{z_1^{(A)2}Q^2}\right ]\,~.
   \label{ka}
   \end{eqnarray}
The A-dependent kinematical variables of the process are
\begin{eqnarray}   y_A = \frac{k_1\, q}{k_1\,
k_e} ~, \quad
 x_A = {x_{Bj} \over z_1^{(A)}}, \quad
z_1^{(A)} = {k_1  q \over m_N \nu}~, \label{adef}
\end{eqnarray}
In Eq.~(\ref{ka})   $\alpha_{em}$ denotes the electromagnetic fine
structure constant and  in   Eq.~(\ref{crosa-1}) the angular
dependence of ${\bf P}_{A-1}$ is provided by $y_A$ and
$z_1^{(A)}$. Note, that since  at high values of $Q^2$ one has
$y\sim y_A$
 (in the Bjorken limit $y = y_A$),  the factor $ \left( \frac{y}{y_A}\right)^2$  in the cross
 section (\ref{crosa-1}) is often   omitted (see e.g. Ref. \cite{klimenko}).

 The relevant nuclear quantity in  Eq.~(\ref{crosa-1}) is
\begin{eqnarray}
&&n^A_0(|\Vec P_{A-1}|) = \nonumber\\
 &&=\frac{1}{2 {J}_A+1}
 \sum_{ {\cal M}_{A},{\cal M}_{A-1}} \left | \int\, d {\bf r}_1^{\prime} e^{i {\bf
P}_{A-1} {\bf r}_1^{\prime}} \langle \Psi_{J_{A-1}, {\cal
M}_{A-1}}^{0}( \{{\bf r}_i^{\prime}\})| \Psi_{J_{A}, {\cal
M}_{A}}^{0}( {\bf r}_1^{\prime}, \{{\bf r}_i^{\prime} \}) \rangle
 \right |^2
\label{dismom}
 \end{eqnarray}
which represents  the  momentum distributions   of the hit nucleon
which was bound in the nucleus with minimum removal energy
$E_{min}=|E_A|-|E_{A-1}|$, $E_A$ and $E_{A-1}$ being the
ground-state energies of nuclei $A$ and $(A-1)$, respectively.
In  Eq. (\ref{dismom})  $\Psi_{J_{A}, {\cal M}_{A}}^{0}$ and
$\Psi_{J_{A-1}, {\cal M}_{A-1}}^{0}$ denote the intrinsic ground
state wave functions of nuclei $A$ and $(A-1)$, respectively, ${\bf
r}_1^{\prime}$ describes the motion of the debris with respect to
the center-of-mass (CM) of $(A-1)$,   and, eventually, $\{{\bf r}_i^{\prime}
\}$ stands for a set of $A-2$ intrinsic variables. The nucleon
momentum distributions generate non trivial nuclear effects in
SIDIS, whose nuclear dependence is also given by the quantities
$y_A$ and $z_1^{(A)}$, which
  differ from the corresponding quantities for a free
  nucleon  ($y^N\equiv y=\nu/{E}_e$ and $z_1^{(N)}=1$)  if the off mass shellness of
  the latter  ($k_1^2\neq m_N^2$ ) generated  by nuclear binding is
  taken into account. Equation~(\ref{crosa-1})
  is valid for  finite values of  $Q^2$, and for $A=2$ it agrees with the expression
  used  in Refs.~\cite{simula,fs,valey}.
  Let us now consider the effects of the FSI. This is due
  to the propagation of the nucleon  debris formed after  $\gamma^*$ absorption
   by a  quark,
   followed  by its hadronization, and by the hadronization of the diquark,\\
    and the interactions
 of the  newly produced  hadrons with the nucleons of $(A-1)$. Calculation of such
 a kind of FSI from first principle represents therefore a very complicated many-body
 problem, so that appropriate  model approaches have to be developed. To this end, one is
 guided by the observation that in the kinematics we are considering
  (i)  the   momentum of the spectator nucleus $(A-1)$ is slow;
  (ii) the relative momentum  between the debris (with momentum  ${\bf p}_X$)
  and   nucleon  $i$ of $(A-1)$ (with momentum ${\bf k}_i$),  is   very large,
  i.e. $ |({\bf p}_X-{\bf k}_i)| \simeq |{\bf q}|\gg |{\bf k}_i |$
  and
   (iii) the momentum transfer in the  interaction between the nucleon debris and $(A-1)$ is
   of the order typical for the scale  of   high-energy elastic $NN$ scattering,
   i.e. much smaller than the  incident momenta  $ {\bf p}_X$.
   As a
  consequence, most of the momentum carried by the virtual photon is
  transferred to the hit quark and  the exact rescattering wave
function can be replaced by its eikonal approximation describing
the propagation of the
 nucleon  debris formed after  $\gamma^*$ absorption by a target quark, followed  by
  hadronization processes and interactions of the newly produced  hadrons with the
   spectator nucleons.
  The series of soft
 interactions between the produced hadrons and the spectators nucleons  can be characterized by an effective cross
 section  $\sigma_{eff}(z,Q^2,x_{Bj})$  depending upon time
 (or the distance $z$ traveled  by the system $X$) \cite{ciokop}.
 The SIDIS cross section which includes FSI will therefore read as follows
 \cite{ciokop,ckk,veronica}
\begin{eqnarray}
   &&\!\!\!\!\!\! \sigma^{A,FSI} (x_{Bj},Q^2,|\Vec P_{A-1}|,y_A,z_1^{(A)})
   \equiv\sigma^{A,FSI}=
  \frac{d\sigma^{A,FSI}}{d x_{Bj} d Q^2  d \Vec   P_{A-1}}=\nonumber\\&&
   =  K^A( x_{Bj},Q^2,y_A,z_1^{(A)}) z_1^{(A)}
   F_2^{N/A}(x_A,Q^2,k_1^2)\, n_0^{A,FSI}(\Vec P_{A-1}),
   \label{crossdist}
   \end{eqnarray}\\

\noindent where $n_0^{A,FSI}(\Vec P_{A-1})$ is the  distorted
momentum distribution of the bound nucleon
\begin{eqnarray}
\hspace*{-1cm} &&n_0^{A,FSI}(\Vec P_{A-1}) = \nonumber\\
\hspace*{-1cm} &&\!\! \!\!  =\frac{1}{2J_A+1}
\!\!\sum_{{\cal M}_A,{\cal M}_{A-1}} \left | \int\, d {\bf r}_1^{\prime}
 e^{i {\bf P}_{A-1} {\bf r}_1^{\prime}} \langle \Psi_{J_{A-1}, {\cal M}_{A-1}}^{0}
( \{{\bf r}_i^{\prime}\}) |S_{FSI}^{XN}({\bf r}_1, \dots ,{\bf r}_A)|
\Psi_{J_{A}, {\cal M}_{A}}^{0}( {\bf r}_1^{\prime}, \{{\bf r}_i^{\prime} \}) \rangle
 \right |^2\!\!\!\!.
 \label{dismomfsi}
 \end{eqnarray}
\noindent Here the  quantity
$S_{FSI}^{XN}$ is the the debris-nucleon eikonal scattering
$S$-matrix
 \begin{equation}
S_{FSI}^{XN}({\bf r}_1,\dots,{\bf r}_A)=
\prod_{i=2}^{A}\bigl[1-\theta(z_i-z_1)\Gamma({\bf b}_1-{\bf b}_i,{
z}_1-{z}_i)\bigr] \label{SG}
\end{equation}
with  the $Q^2$- and $x_{Bj}$-dependent profile function being
\beq \Gamma^{XN}({{\bf b}_{1i}},z_{1i})\,
=\,\frac{(1-i\,\alpha)\,\, \sigma_{eff}(z_{1i}, Q^2,x_{Bj})}
{4\,\pi\,b_0^2}\,\exp \left[-\frac{{\bf
b}_{1i}^{2}}{2\,b_0^2}\right],
 \label{eikonal}
\eeq
\noindent  where  ${\bf r}_{1i}=\{{\bf b}_{1i}, {\bf z}_{1i}\}$, with ${\bf z}_{1i} ={\bf z}_{1}-{\bf
z}_{i}$  and ${\bf b}_{1i}={\bf b}_{1}-{\bf b}_{i}$. It can be seen that, unlike
the standard Glauber eikonal approximation~\cite{glau2}, the profile function
$\Gamma^{XN}$ depends not only upon the  transverse relative  separation
but also upon the longitudinal separation $z_{1,i}={z}_1-{z}_i$ due to the $z$- (or time)
 dependence of the effective cross section $\sigma_{eff}(z_{1i})$ and the
  $\theta$-function, $\theta(z_i-z_1)$.
As already mentioned, the effective cross section
$\sigma_{eff}(z_{1i})$ also depends on the total energy of the
debris, $W_X^2\equiv P_X^2$; if the energy is not high enough, the
hadronization procedure can terminate inside the nucleus $(A-1)$,
after which  the number of produced hadrons and the  cross section
$\sigma_{eff}(z_{1i},x_{Bj},Q^2) \equiv \sigma_{eff}(z)$ remain
constant \cite{veronica}
\subsection{The effective debris-nucleon cross section}
   As already  pointed out, although the profile function given by  Eq.~(\ref{eikonal}) resembles
the usual Glauber form, it contains an important difference, in
that it  depends also upon the longitudinal separation
$z_{1i}={z}_1-{z}_i$ due to the $z$- (or time) dependence of the
effective cross section $\sigma_{eff}(z)$, which describes the
interaction of the debris of the so called {\it active  nucleon}
"1" with the spectator nucleon "i".
 The effective cross section
$\sigma_{eff}(z)$ has been derived in detail in Ref.
\cite{ciokop}. It has  already been used in the description of
SIDIS off nuclei  in Refs. \cite{ckk,veronica} and has been shown
\cite{grey} to  provide a good description of grey track
production in muon-nucleus DIS at
 high energies~\cite{adams}. Therefore, it only suffices   to recall here  that at the given point
 $z$, $\sigma_{eff}(z)$
 consists of   a sum of the  nucleon-nucleon and   pion-nucleon total cross sections,
 with the former describing
 the hadronization of the diquark into a nucleon
 and the latter increasing with $z$
 like the multiplicity of pions produced by the
breaking of the color string and by gluon radiation, respectively,
namely   $\sigma_{eff}(z)$ has the following form:
$\sigma_{eff}(z)\,=\,\sigma_{tot}^{NN}\,+\,\sigma_{tot}^{\pi N}\,
\big[\,n_{M}(z)\,+\,n_{G}(z)\,\big]$,  where the $Q^2$- and
$x_{Bj}$-dependent quantities $n_{M}(z)$ and $n_{G}(z)$,  are the
pion multiplicities due to the breaking of the color string and to
gluon radiation; their explicit forms are given in Ref.
\cite{ciokop}.
  Let us stress  that hadronization  is basically a QCD
 nonperturbative  process, and, consequently, any   experimental information on its
 effects on the SIDIS process we are considering would be a rather valuable one;
 since it has been shown in Ref.~\cite{ckk} that in the kinematical range where
 FSI effects  are relevant the process  is essentially governed by the  hadronization
 cross section, this opens a new and important  aspect of these reactions, namely
 the possibility, through them, to investigate  hadronization mechanisms
  by choosing a proper kinematics where FSI effects are maximized.

\begin{figure}[!ht]
\vskip 10cm
\centerline{
\includegraphics[scale=0.55 ,angle=0]{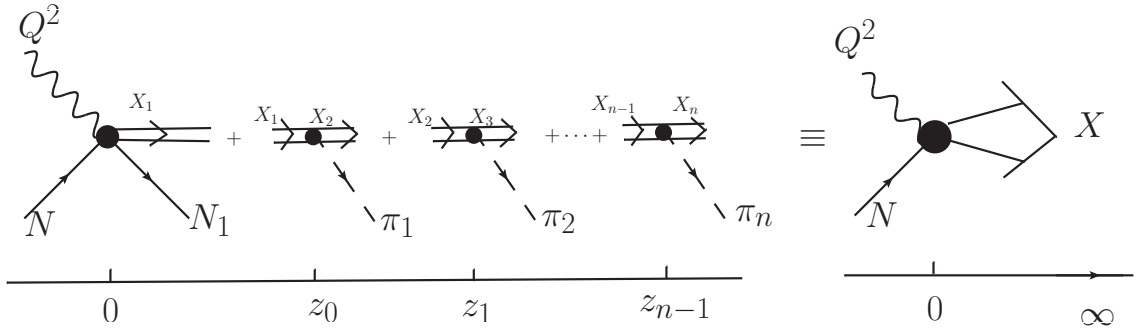}}
\caption{Schematic representation of pion and nucleon $N_1$ production  by quark and diquark hadronization  leading to the
 FSI in the SIDIS process $A(e,e'(A-1))X$ .}
\label{Fig2}
\end{figure}

In Ref. \cite{ciokop} $\sigma_{eff}(z)$ was obtained in the limit
of very high energies. In this paper, as in Ref. \cite{veronica},
we generalize the results of Ref. \cite{ciokop} to  lower energies
(e.g. JLab ones)  by the following procedure.  According to the
hadronization model of Ref.~\cite{koppred}, the process of pion
production on a nucleon after  $ \gamma^*$ absorption  by a quark
can schematically be represented as in
 Fig. \ref{Fig2}: at the interaction point  a color string, denoted  $X_1$, and
 a nucleon $N_1$, arising from target fragmentation,
  are formed; the  color  string
propagates  and gluon radiation begins.  The first "pion" is
created  at $z_0 \simeq 0.6$ by the breaking of the color string
and  pion production continues until
it stops at a maximum value of $z =z_{max}$,  when energy conservation does not allow
further
 "pions"
to be created,  and  the number of pions remains constant;  we
obtain
\begin{equation}
 z_{max}=\frac {E_{loss}^{max}}{\kappa_{str}+\kappa_{gl}}=
 \xi
 \frac{E_X -E_{N_1}}{\kappa_{str}+\kappa_{gl}}
 \label{nove}
\end{equation}
where (see Ref. \cite{koppred}) $\kappa_{gl}=2/(3\,\pi)
\alpha_{s}(Q^2-\lambda^2)$  (with  $\lambda\approx 0.65\,GeV$) and
$\kappa_{str} = 0.2 \, GeV^2$ represent the energy loss
 ($\kappa = -dE/dz$) of the leading hadronizing quark
 due to the string breaking and gluon radiation, respectively;
 in  Eq. (\ref{nove}) $E_{loss}^{max}$ is the maximum energy
 loss, which can be expressed in terms of the energy of the nucleon debris $E_X$ and the
energy $E_{N_1}$ of the nucleon created by
 target fragmentation at the interaction point.  The maximum energy loss depends  upon the kinematics of the
 process, and within the kinematics we have considered, it turns
 out that
 $\xi=0.55$. It should be pointed out that once the total effective cross section $\sigma_{eff}(z)$ has been obtained,
  the elastic slope $b_0$ and
 the ratio $\alpha$ of the real to the imaginary parts of the elastic amplitude remain to be determined. This does not represent
 a problem at very high energies and for medium and  heavy nuclei, as considered in Ref. \cite{ciokop},
 since in this case $\alpha \rightarrow 0$ and, within
 the optical limit, the cross section will only depend upon
the convolution between the effective cross section
$\sigma_{eff}(z'-z)$ and the nuclear density $\rho({\Vec b}, z')$
i.e. the quantity $S({\Vec b}, z) =\int d\,z'\,\, \rho({\Vec b},
z')\sigma_{eff}(z'-z)$. At lower energies $\alpha$ and $b_0$
appear explicitly  in the calculations. Their choice will be
discussed in the next Section.

\section{PROCESS $^2H(e,e'p)X$}
\subsection{Details of calculations}
Within the spectator mechanism, $\gamma^*$ interacts with a quark
of the neutron and  the spectator  proton recoils   and is
detected with momentum
 ${\bf P}_{A-1} \equiv{\bf p}_p$, with ${\bf p}_p=-{\bf k}_1$  in PWIA,
  and ${\bf p}_p \neq {\bf k}_1$ when FSI is
 considered ({\it cf.} Fig. \ref{Fig1})
 (note that the detected nucleon momentum ${\bf p}_p$ is denoted ${\bf p}_2$ in Ref. \cite{veronica} and ${\bf p}_s$
 in Refs. \cite{klimenko}, \cite{kuhn}, \cite{fs} and \cite{simula}). We have calculated
 the process $^2H(e,e'p)X$
 at the kinematics of the recent Jlab experiment
\cite{klimenko,kuhn} both in PWIA (Eq.~(\ref{crosa-1})), and taking
FSI into account (Eq.~(\ref{crossdist})). We have used deuteron wave
functions generated by realistic $NN$ potenlias, in particular the
$AV18$ interaction~\cite{AV18}.  For the nucleon deep inelastic
structure function $F_2$ we have used the parametrization
 from Ref.~\cite{effe2}  with the nucleon off-mass shell within the x-rescaling
model, i. e. by using $x_A=x_{Bj}/z_1^{(A)}$, where
$z_1^{(A)}=k_1\cdot q/(m_p\nu)$ with $k_1^0
=M_D-\sqrt{{m_p}^2+{{\bf p}_p}^2}$ (in what follows all
quantities, e.g. mass, momentum, cross section, etc., pertaining
to  $^2H$ will be labeled by a capital $D$.)
 As for  the quantities appearing in the
profile function (\ref{eikonal}) we have used the following
procedure, which is  appropriate for the kinematics we have
considered (see next Subsection): $\sigma_{eff}$ has been
calculated as explained in Section II with values $\sigma_{NN} =
40 \,mb$ and $\sigma_{\pi N} = 30\, mb$, and $\alpha$ and $b_0$
taken from world data on $\pi N$ scattering, since  the underlying
FSI mechanism is described by the $\pi N$ cross section and  the
pion multiplicities. The comparison of the results of our
calculations, which contain no adjustable parameters, are
presented in the next subsection and compared with available
experimental data.
\subsection{Comparison with experimental data and the effect of FSI}
Experimental data on the process $^2H(e,e'p)X$ have  recently been
obtained at Jlab~\cite{klimenko,kuhn} in the following kinematical
regions: beam energy $E_e = 5.75\,\, GeV$, four-momentum transfer
$1.2 \,\,(GeV/c)^2 \lesssim Q^2 \lesssim 5.0 \,\,(GeV/c)^2$,
recoiling proton momentum $0.28 \,\,GeV/c \lesssim|{\bf
p}_{p}|\leq 0.7\,\, GeV/c$, proton emission angle $-0.8  \leq
\cos\theta_{\bf p} \leq 0.7$ $(\theta_{\widehat {{\bf
p}_p\cdot{\bf q}}} \equiv \theta_{\bf p})$, invariant mass of the
produced hadronic state $1.1 \,\,GeV\,\,\leq W_X\leq \,\,2.7 \,\,
GeV$, with $W_X^2={(k_1+q)}^2={(P_D- p_p+q)}^2$. The data have
been plotted in terms of the reduced cross section
\beq \sigma^{red}(x_{Bj},Q^2,{\bf p}_p) = \frac{1}{K^A(
x_{Bj},Q^2,y_A,z_1^{(A)}) }\,\left(
\frac{y}{y_D}\right)^2\frac{d\sigma ^{D,exp}}{d x_{Bj} d Q^2  d
\Vec p_p} \label{reduced} \eeq
which, within our approach,
would be
 \beq
\sigma^{red}(x_{Bj},Q^2,{\bf p}_p) = \left( \frac{y}{y_D}\right)^2
z_1^{(D)} F_2^{N/D}(x_D,Q^2,k_1^2)\, n_0^{D,FSI}({\bf p}_{p})
\label{reducedour} \eeq
\noindent in agreement with the experimental definition  of Ref. \cite{klimenko}.
 A comparison between theoretical
calculations and the experimental data plotted {\it vs} $\cos \theta_{{\bf p}}$
at fixed values of $Q^2$, $W_X$ and  $|{\bf p}_p|$,
is presented in Fig.~\ref{Fig3}, which clearly shows that:
 i) apart from the very backward emission, the
experimental data are dominated by the FSI; ii) our model of FSI
provides a satisfactory description of the experimental data in
the backward direction and also around $\theta_{{\bf p}} \simeq
90^{\circ}$ (a comparison of theoretical results and experimental
data in the full range of kinematics of Ref.~\cite{klimenko, kuhn}
will be presented elsewhere); and (iii) in the forward direction
($\theta_{{\bf p}} \lesssim 80^{\circ}$) the spectator mechanism
  fails to reproduce the data and it is clear that other
production mechanisms   are playing a role in this region. For such a reason
 in what follows we will consider  the region ($\theta_{{\bf p}}
\gtrsim 80^{\circ}$) where useful information on both the
hadronization mechanism and the EMC effect can in principle be
obtained, provided the data are analyzed in the proper way,
getting rid of  EMC effects, in the former case, and of nuclear
effects, in the latter case. This problem will be clarified in the
next Section on the example of the process $^3He(e,e'\,{d})X$.

\section{PROCESS $^3He(e,e'\,D)X$}
\subsection{Details of calculations}
In the process $^3He(e,e'\,d)X$ the virtual photon
 $\gamma^*$ interacts with a quark of the proton and the spectator deuteron
recoils and is detected with momentum
 ${\bf P}_{A-1} \equiv {\bf P}_D$, with ${\bf P}_D=-{\bf k}_1$  in PWIA and ${\bf P}_D \neq {\bf k}_1$
 when FSI is considered ({\it cf.} Fig.~\ref{Fig1}). We  considered
 the process at  kinematics  similar  to the ones of the $12-GeV$ upgraded Jlab.
 Nuclear structure effects were taken care
 of  within a full self consistent and realistic approach based upon the
 deuteron and ${^3}He$ wave functions obtained from an exact solution of
 the Schr\"odinger equation corresponding to the $AV18$ $NN$ interaction; in particular,
 the three-nucleon
 wave functions correspond to those of Ref. \cite{rosati}.
For the
nucleon structure function $F_2^{N/A}$
 we  used the  parametrization from Ref.~\cite{effe2},   with the nucleon off-mass shell
 within the x-rescaling model, i.e. $x_A=x_{Bj}/z_1^{(A)}$ where   $z_1^{(A)}=k_1\cdot
q/(m_N\nu)$ with  $k_1^0 =M_3-\sqrt{{M_D}^2+{{\bf p}_D}^2}$.
 Within such a framework, we demonstrate in the  next sections how to tag the
hadronization mechanism and the EMC effect, i.e. how to obtain
information on (i) hadronization mechanisms, free from possible
contaminations of unknown EMC effects, and (ii) the EMC effect,
free from possible contamination of unknown nuclear structure
effects.

\subsection{Tagging the hadronization mechanisms}
In  Fig.~\ref{Fig4} we show the cross section of the process
$^3He(e,e'\,d)X$ calculated by Eqs.~(\ref{crosa-1})
and~(\ref{crossdist}) at two different values of the deuteron
emission angle, corresponding to parallel ($\theta_{\widehat{\Vec
P_{D}  \Vec q}} \equiv \theta_{D}=180^{o}$) and perpendicular one
($\theta_{\widehat{\Vec P_{D} \Vec q}} \equiv \theta_{D}=90^{o}$)
kinematics, respectively. It can be seen that, as in the case of
the process $^2H(e,e'p)X$, FSI  increases with the momentum of the
detected deuteron  and is particular relevant in perpendicular
kinematics,  which is the region one has to consider to obtain
 information about hadronization mechanisms. To this end, in order to
 minimize possible contaminations from the poor knowledge of the neutron structure function,
 the
 ratio of the cross sections for two different nuclei $A$ and
$A^{\prime}$ measured at the same value of $x_{Bj}$ should be
considered, since, within our approach, one has
 \begin{eqnarray}
   R^{exp}(x_{Bj},Q^2,|\Vec P_{A-1}|,z_1^{(A)},z_1^{(A')},y_A ,y_{A'})
 &=& \frac{\sigma^{A,exp} ( x_{Bj},Q^2,|\Vec P_{A-1}|,z_1^{(A)},y_A ) }
  {\sigma^{A',exp} ( x_{Bj},Q^2,|\Vec P_{A-1}|,z_1^{(A')},y_{A'} ) }
  \rightarrow \nonumber\\
 &\rightarrow&\frac{{n_0^{(A,FSI)}(\Vec P_{A-1})}}{{n_0^{(A',FSI)} (\Vec
 P_{A-1})}} \equiv R(A,A',\Vec P_{A-1})
   \label{ratioa-1}
   \end{eqnarray}
where the last step, as stressed in Ref.~\cite{scopetta}, is not valid exactly,
for the  factors  $K( x_{Bj},Q^2,y_A,z_1^{(A)})$ and   $z_1^{(A)}F_2^{N/A}(x_D,Q^2,k_1^2)$ in Eq. (\ref{crosa-1})
 depends upon $A$ via $y_A$ and $z_1^{(A)}$. However, as discussed in detail in Ref. \cite{scopetta}
  (see Fig. 3 there and the discussion after Eq. (32)), at high
 values of $Q^2$ the factor $K^A$ differ only by a few percent from the
 free nucleon value $K^N$,  i.e. becomes practically A-independent ($K^A=K^N$ in the Bjorken limit); as for the A-dependence of the ratio
 $z_1^{(A)}F_2^{N/A}/z_1^{(A')}F_2^{N/A'}$, it can be of the order of the EMC effect and in Ref. \cite{scopetta} has been numerically estimated
 to be at maximum of 5 \%
 (note that in the Bjorken limit and within  a light cone approach, the ratio is
 exactly unity, being the nucleons on shell);
 at the same time,
  in the low nucleon momentum
 region we are considering, the momentum distributions of light nuclei
 may differ up to an order of magnitude (see Fig.2 of
Ref. \cite{scopetta}); it is clear therefore  that
 the $|{\Vec P}_{A-1}|$-dependence of the
ratio (\ref{ratioa-1}) is  governed  by the ratio of the momentum
distributions and  any reasonably expected  A-dependence of
$F_2^{N/A}(x_A,Q^2,k_1^2)$ through $x_A$ will not affect it.  The
ratio for $A=2$ and $A'=3$ is shown in Fig.~\ref{Fig5}. It can be
seen, that at low values of the detected momentum,  FSI plays only
a minor role both in parallel and perpendicular kinematics. In
this respect, we would like to stress once again that whereas the
momentum dependence of the cross section is generated both by the
momentum dependence of the distorted momentum distributions, and
by a possible momentum dependence of the nucleon structure
functions (see next Section), the momentum dependence of the ratio
(\ref{ratioa-1})  is only governed by the distorted momentum
distribution; since the low momentum part of the momentum
distribution
 is very well known for A=2 and A=3 systems (as well as for heavier nuclei),
the experimental observation of a  ratio    similar to the
one shown in Fig.~\ref{Fig5}, would provide strong evidence of the
correctness of the spectator mechanism and of the FSI model; at the
same time, the observation of strong deviations from the
prediction shown in Fig. \ref{Fig5}, would provide evidence of
reaction mechanism and/or FSI effects which are missing in our
model. Experiments on  heavier nuclei, particularly at
perpendicular kinematics and $|{\bf p}_{A-1}| \simeq 0.2 \div 0.4
\,\,GeV/c$ ({\it cf.} Fig. \ref{Fig4}),  where the effects of FSI are
expected to be more relevant \cite{ciokop}, would be extremely
useful to clarify the mechanism of the FSI.

\subsection{Tagging the EMC effect}
In order to tag the EMC effect, i.e. if, how,   and to what extent
the nucleon structure function in the medium differs from the free
structure function, one has to get rid  of  the effects due to the
distorted nucleon momentum distributions and other nuclear
structure effects, i.e. one has to consider a quantity which would
depend only upon $F_2^{N/A}(x_{A},Q^2,k_1^2)$. This can be
achieved by considering the ratio of the cross sections on nucleus
$A$ measured at two different values of the Bjorken scaling
variable $x_{Bj}$ and $x_{Bj}^{\prime}$, leaving unchanged all
other quantities in the two cross sections,  i.e. the ratio
\begin{eqnarray}
 R^{exp}(x_{Bj}, x_{Bj}^{\prime}, Q^2,|\Vec P_{A-1}|,z_1^{(A)},y_A )&=&
  \frac{\sigma^{A,exp} ( x_{Bj},Q^2,|\Vec P_{A-1}|,z_1^{(A)},y_A ) }
             {\sigma^{A,exp} (  x_{Bj}^{\prime},Q^2,|\Vec P_{A-1}|,z_1^{(A)'},y'_{A} ) }
             \rightarrow \nonumber\\
&\rightarrow&\frac{F_2^{N/A}(x_A,Q^2,k_1^2)}
{F_2^{N/A}(x_{A}^{\prime},Q^2,k_1^2)}  \equiv
   R(x_{Bj}, x_{Bj}^{\prime}, |\Vec P_{A-1}|)
   \label{ratioa}
\end{eqnarray}

We  considered  the quantity  (\ref{ratioa}) calculated in the
following kinematical range:
  $2\,\,\lesssim W_X^2 \lesssim \,\, 10\,\,GeV{^2}$ and $Q^2= 8\ (GeV/c)^2$.
 At   each  value of $W_X$   we changed  $|{\bf P}_D|$ from zero to
 $|{\bf P}_D| = 0.5 \,\,GeV/c$, obtaining  for different values of $|{\bf P}_D|$
  different values of $x_{Bj}$. To minimize the effects of FSI, the
 angle $\theta_{\widehat {{\bf P}_D\cdot{\bf q}}}$ was chosen  in  the  backward
 direction, $\theta_{\widehat {{\bf P}_D\cdot{\bf q}}}\sim 145^o $ ({\it cf.} Fig.~\ref{Fig4}).
Within such a kinematics the effective cross section $\sigma_{eff}(z_{1i},x_{Bj},Q^2)$
is the same for different values of $W_X$  and,  correspondingly,
the  distorted momentum distributions $n_0^{A,FSI}$ will depend only upon  $|{\bf P}_D|$
and cancel in the ratio (\ref{ratioa}).
By this way, all nuclear structure effect, except possible effects of in-medium deformations
of the nucleon structure function $F_2^{N/A}$, are eliminated and one is left with a ratio which depends
only upon the nucleon structure function $F_2^{N/A}$.
Calculations  have been performed using three  different  structure functions
$F_2^{N/A}(x_{A},Q^2,k_1^2)$, namely:
\begin{enumerate}
\item
the free nucleon structure function from Ref.~\cite{effe2}, exhibiting no EMC effects;
\item
the nucleon structure function pertaining to the x-rescaling model  with
the nucleon off-mass shell, i.e. $F_2^{N/A}(x_{A},Q^2,k_1^2)
\rightarrow F_2^{N}(x_{A},Q^2) =F_2^{N} (\frac{
x_{Bj}}{z_1^A},Q^2)$,
 where   $z_1^A=k_1\cdot q/(m_p\nu)$ with $k_1^0 =M_A-\sqrt{{M_{A-1}^*}^2+{{\bf k}_1}^2}$;
\item
the structure function from Ref.~\cite{CKFS},  which assumes that
 the reduction of nucleon point like configurations (PLC)  in the medium (see Ref.\cite{fs}) depends upon  the nucleon virtuality:
\begin{equation}
F_2^{N/A}(x_{A},Q^2,k_1^2) \rightarrow F_2^{N/A}
 \left ( { x_{Bj}/z_1^N, Q^2}  \right ) \delta_A(x_{Bj},v(|{\bf k_1}|,E)),
\label{del}
 \end{equation}
  where
$z_1^N=(m_N +|{\bf P}_D|\cos\theta_{D})/m_N$. Here the reduction
of  the PLC is given by the quantity  $\delta_A(x_{Bj},v({\bf
k},E))$, which depends upon the nucleon virtuality (see
\cite{CKFS}):
\begin{eqnarray}
\hspace{-0.5cm} v(|{\bf k}_1|, E)=
\left(M_A -\sqrt{(M_{A}- m_N+E)^{2}+{\bf k}_1^{2}}\right)^2-{\bf k}_1^2
- m_N^2.
\label{virtuality}
\end{eqnarray}
\end{enumerate}

It should be stressed that the two medium-dependent structure
functions provide similar results for the inclusive cross section
and that our aim is to answer the question as to
whether the  SIDIS experiment we are proposing could discriminate
between the two models. The results of calculations corresponding
to the kinematics $E_e = 12\,\,GeV$,  $Q^2\,\,=8\,\,{(GeV/c)}^2$,
 $\theta_D =145^{o}$,  $x_{Bj} = 0.45 $, $x_{Bj}^{\prime}=
0.35$ are presented in Fig.~\ref{Fig6}. It can be seen that the
discrimination between different models of the virtuality
dependence of $F_2^{N/A}(x_{A},Q^2,k_1^2)$ can indeed be achieved
by a measurement of the ratio~(\ref{ratioa}); as a matter of fact
at $|{\bf P}_D| \simeq 0.4\,\, GeV/c$ the two structure functions
differ by about $40\, \%$.

\section{SUMMARY AND CONCLUSIONS}

In this paper we  considered the SIDIS process $A(e,e'(A-1))X$ on
complex nuclei proposed in Ref.~\cite{scopetta} within the
spectator model and the PWIA,  and extended in Ref.~\cite{ciokop}
by the inclusion of the FSI between the hadronizing debris and the
nucleons of the detected nucleus $(A-1)$. We focused  on $^2H$ and
$^3He$ targets and extended the  treatment of FSI, by considering
it not only at very high energies, as originally done in
Ref.~\cite{ciokop}, but also at lower energies like the ones of
Jlab. The  reason for considering $^2H$ and $^3He$ is twofold:
 (i)nuclear effects can  accurately be calculated, since realistic wave functions
resulting from the exact solution of the Schr\"odinger equation
can be used, and (ii) experimental data for deuteron targets  have
recently been  obtained \cite{klimenko}.
 The results of our calculations for
  the process $^2H(e,e'p)X$ show that
the experimental data  can be well  reproduced in the kinematics
when the proton is emitted mainly backward in the range $70^{o}
\lesssim \theta_{\bf p} \lesssim 145^{o}$,
 with the effects of  FSI interaction being very small in the very backward
 direction and dominating the cross section around
$\theta_{\bf p} \simeq 90^{o}$. It is very gratifying to see that the experimental
 data can be reproduced in
a wide kinematical region, which make us confident of the
correctness of the spectator model and the treatment of the FSI
between the hadron debris and the detected proton. At emission
angles $\theta_{\bf p} \lesssim 80^{o}$, the number of detected
protons is much higher than our predictions, which is clear
evidence of the presence of production mechanisms different from
the spectator one.  Among possible mechanisms  leading to forward
proton production, target and/or current fragmentation should be
the first processes to be taken into account.
 The first one has been analyzed in Ref.
\cite{veronica}, finding that it contributes only forward and at
proton momenta much higher than the ones typical of the Jlab
kinematics we have considered. The contribution from current
fragmentation effects is under investigation.  It is clear that
the SIDIS process on heavier nuclei,
 with detection of a complex nucleus $(A-1)$ (e.g $^2H$, $^3He$, etc.) would be
  extremely useful,
since  the only mechanism for producing a recoiling $(A-1)$
nucleus would be the spectator mechanism. These experiments would
be very useful  in clarifying  the origin of the discrepancy
between theory and experiment we  found in the forward hemisphere
in the process $^2H(e,e'p)X$. Also, as stressed in Ref.
\cite{ciokop}, they would be very useful in studying the early
stage of hadronization at short formation times
 without being affected by cascading processes, unlike the DIS inclusive hadron
production $A(e,e'h)X$ where most hadrons with small momentum
originate from cascading of more energetic  particles.
  We have
illustrated how by measuring the reduced cross section on two
different nuclei at the same value of the detected momentum, the
validity of the spectator mechanism and information on the
survival probabilities of the spectator nuclei, i.e. on the
hadronization mechanism, could be obtained; moreover,
 by measuring the cross section on the same nucleus,
 but at two different values of $x_{Bj}$, the EMC effect could be tagged.
Experiments of the type we have discussed, e.g. $^2H(e,e'p)X$,
$^3He(e,e'\,d)X$,  $^4He(e,e'\ ^3H)X$, $^4He(e,e'\ ^3He)X$
 would be extremely useful, and it is gratifying to see that such
 experiments are being planned
 thanks to the development of
proper recoil detectors ~\cite{proposals}. We would like to
mention that we  performed our calculations at the upgraded Jlab
kinematics, but, as suggested in Ref.~\cite{scopetta},
  the SIDIS process we are
proposing  could in principle  be investigated at an electron-ion collider,
 where higher values of  $Q^2$ and a wider interval of $x_{Bj}$  could be reached,
  which would make the basic assumption of the spectator model,
 {\it viz}, the factorization assumption,  even more reliable.
 The analysis of relevant kinematics and cross sections are
 underway and will be reported elsewhere.

\section*{ACKNOWLEDGMENTS}
\begin{figure}[t]
\vskip 6cm


 \includegraphics[width=0.47\textwidth,height=0.35\textheight]{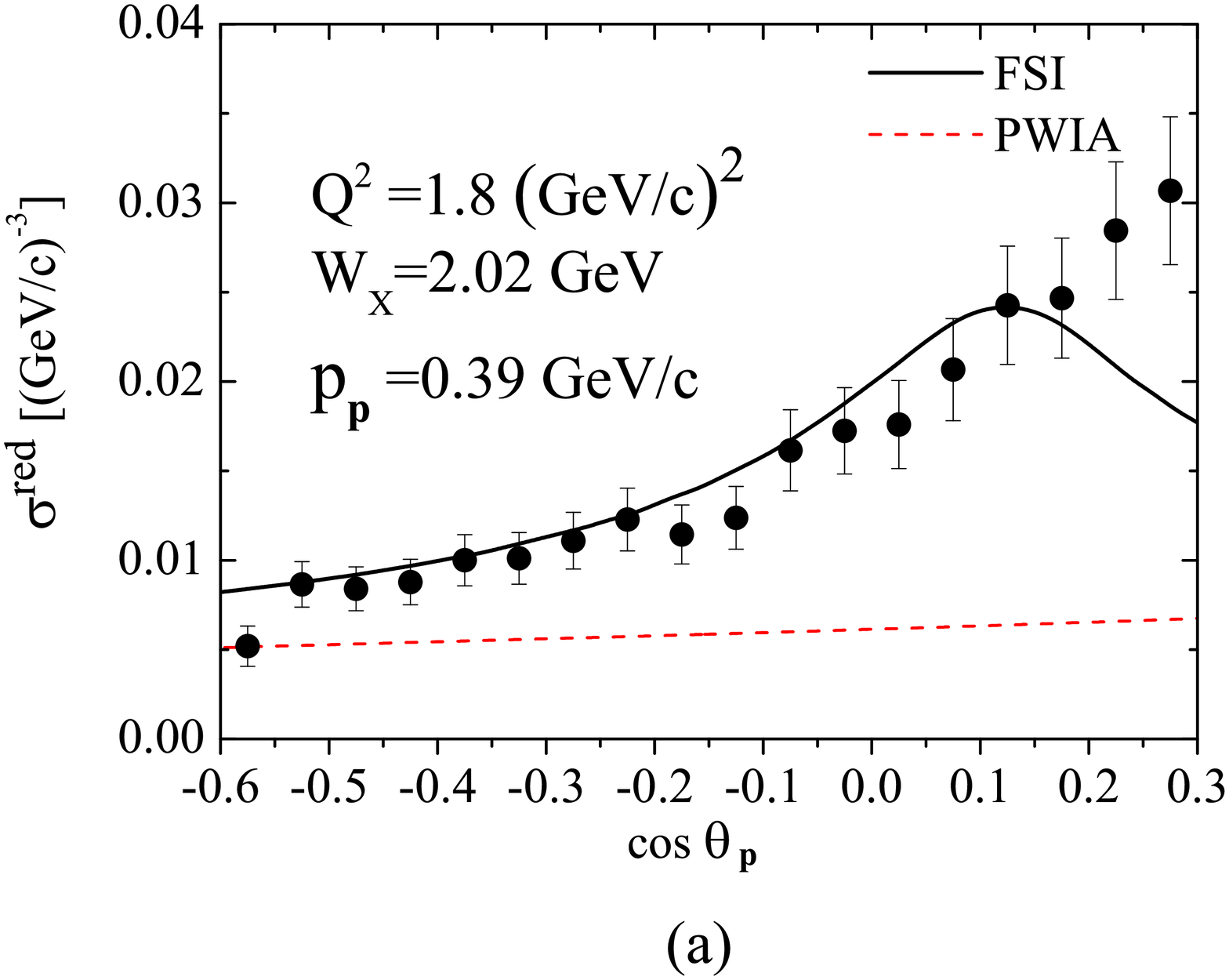}
 \includegraphics[width=0.47\textwidth,height=0.35\textheight]{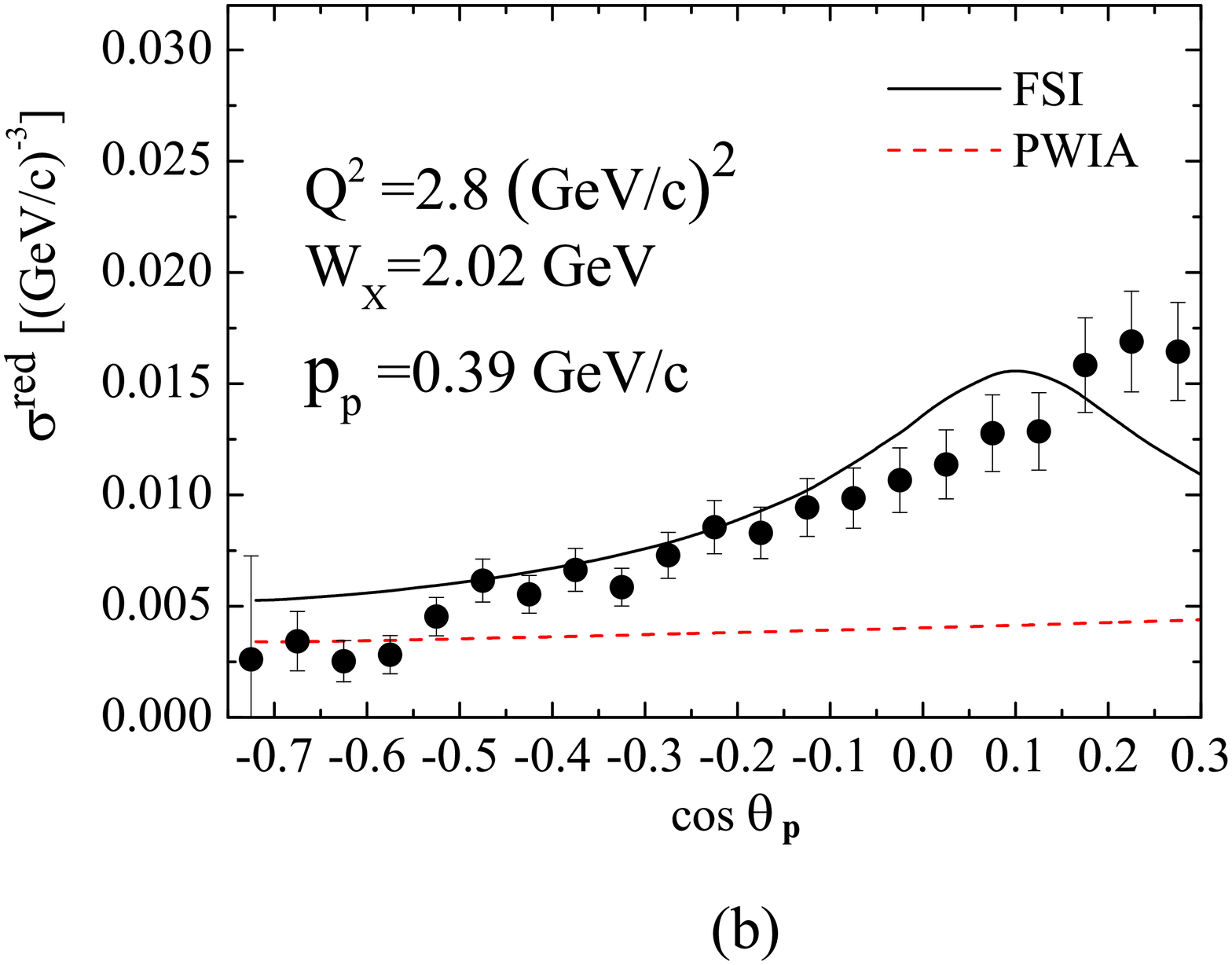}\vspace*{-10mm}
 \includegraphics[width=0.47\textwidth,height=0.35\textheight]{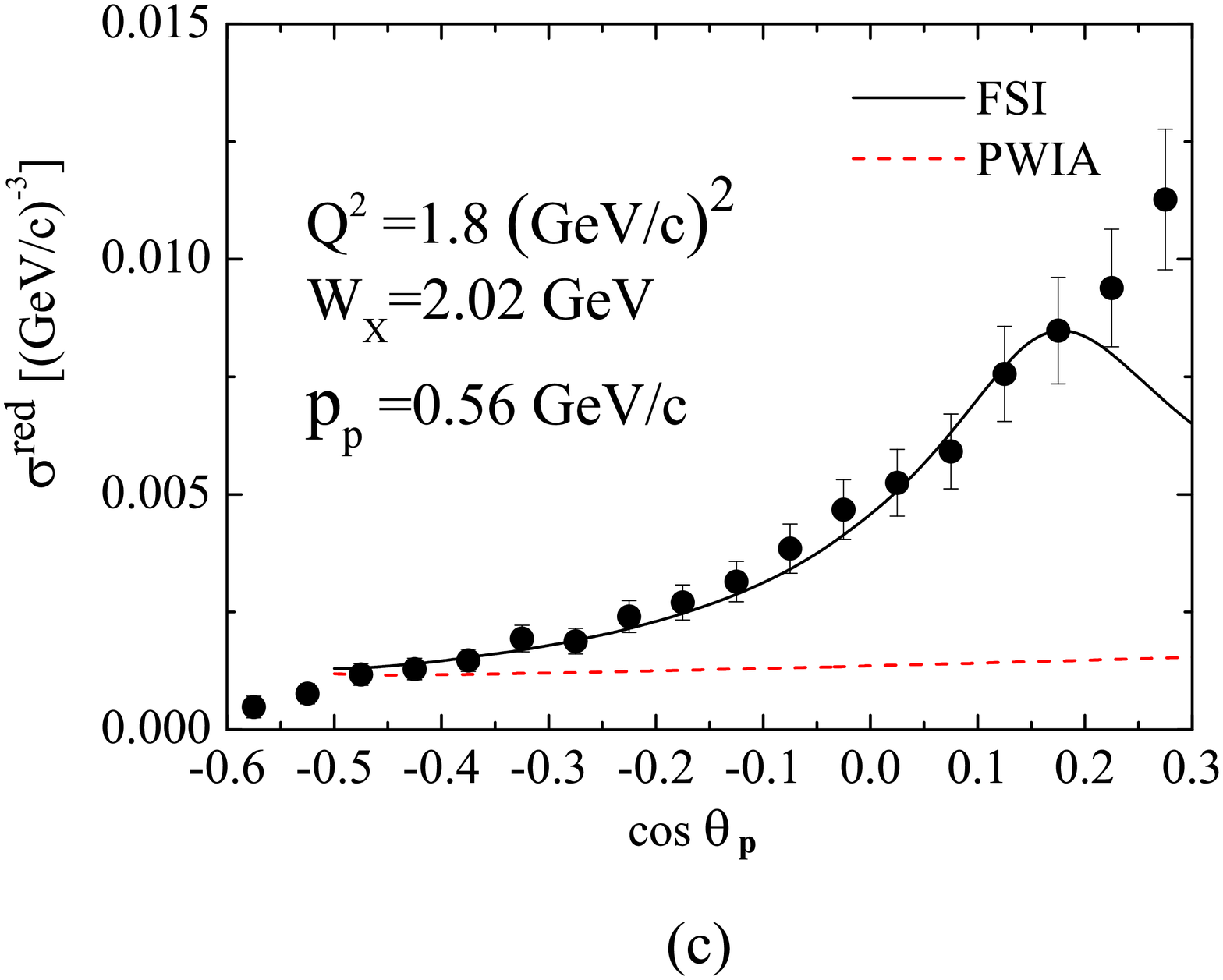}
 \includegraphics[width=0.47\textwidth,height=0.35\textheight]{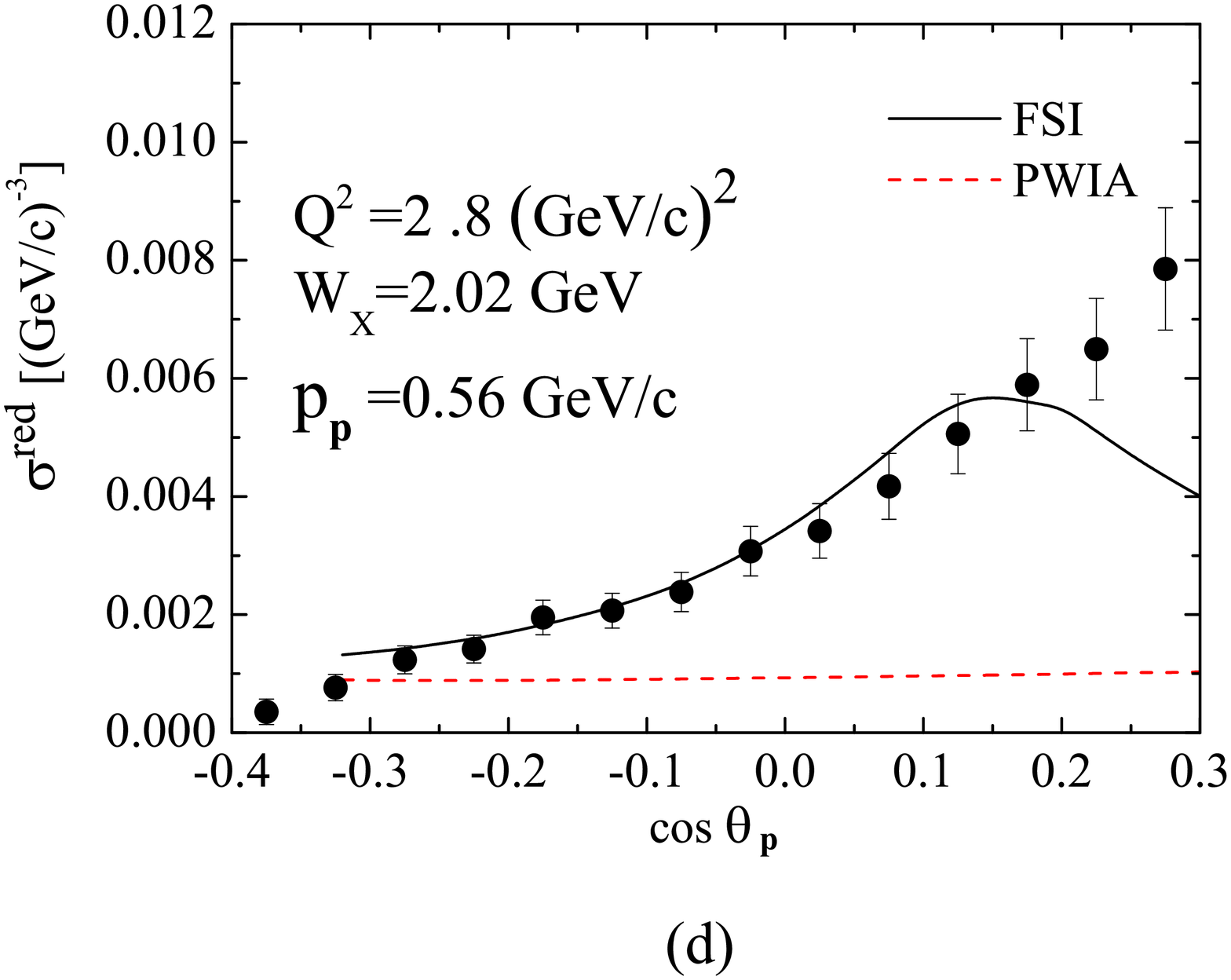}
\caption{(Color online)The theoretical reduced cross section, Eq.
(\ref{reducedour}), {\it vs} $cos \theta_{\widehat{{\bf q} {\bf
p}_p}}$  ($\theta_{\widehat{{\bf q} {\bf p}_p}} \equiv \theta_p$)
compared with the experimental data of Ref. \cite{klimenko, kuhn}.
Each Figure shows the reduced cross section calculated at fixed
values of the four-momentum transfer $Q^2$, the invariant mass
$W_X$ of the hadronic state $X$, and the momentum $|{\bf p}_p|
\equiv p_p$ of the detected proton. The error bars represent the
sum in quadrature of statistical and systematic errors given in
Refs. \cite{klimenko, kuhn}}
\label{Fig3}
\end{figure}

\begin{figure}[!hcp]
\centerline{\hspace{0.2cm}
\epsfysize=8.5cm\epsfbox{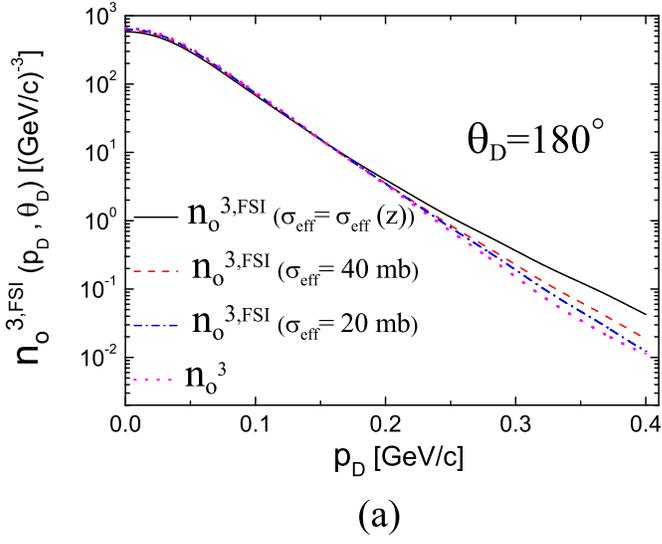}\hspace{0.05cm}
    \epsfysize=8.5cm\epsfbox{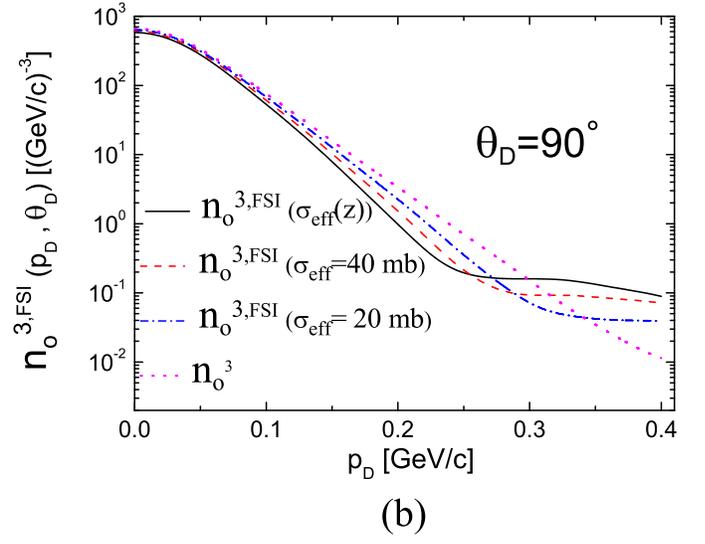}\hspace{0.05cm}}
\caption{The distorted momentum distribution $n_0^{3,FSI}({\bf
P}_{A-1})$ (Eq. (\ref{dismomfsi}) with ${\bf P}_{A-1} \equiv {\bf
p}_{D}$) in the process $^3He(e,e'd)X$ in parallel (a)
and perpendicular (b) kinematics calculated with
different effective debris-nucleon cross sections in Eq. (\ref{eikonal}): the effective  debris-nucleon
cross section $\sigma_{eff}(z) \equiv \sigma_{eff}(z,Q^2,x_{Bj})$  (full line) and two constant cross sections (dashed and dot-dashed lines).
Also shown
(dotted line) is the momentum distribution $n_0^{3}(|{\bf
P}_{A-1}|)$ (Eq. (\ref{dismom})). Caluclations have been performed
at the following kinematics:  $E_e=12 \ GeV$, $Q^2=6 \ GeV^2/c^2$ and $W_X^2=5.8\
GeV^2$.}
    \label{Fig4}
\end{figure}

\begin{figure}[!hcp]
\vskip 6cm
\includegraphics[scale=1.2]{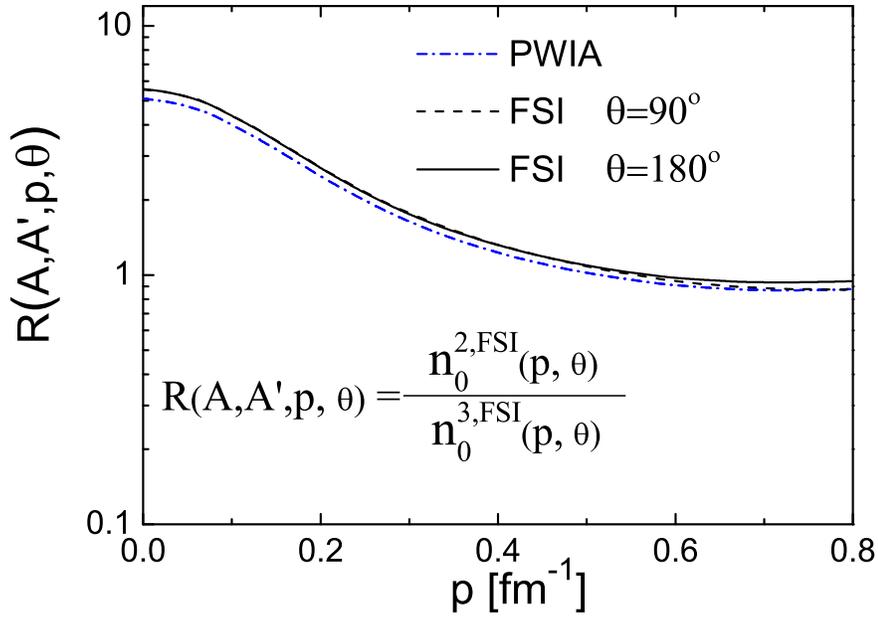}
\caption{The ratio given by Eq.~(\ref{ratioa-1}) with  $A ={^2}H$  and $A'={^3}He$,
corresponding to the
processes $^2H(e,e'p)X$ and ${^3}H(e,e'd)X$ in parallel
($\theta_p=\theta_D \equiv \theta =180^{o}$) and perpendicular
 ($\theta_p=\theta_D \equiv \theta =90^o$)  kinematics, respectively. Protons and deuterons are
detected with the same value of the momentum $|{\bf p}_p|=|{\bf p}_D|\equiv p$.}
\label{Fig5}
\end{figure}

\begin{figure}[!hcp]
\includegraphics[scale=0.50]{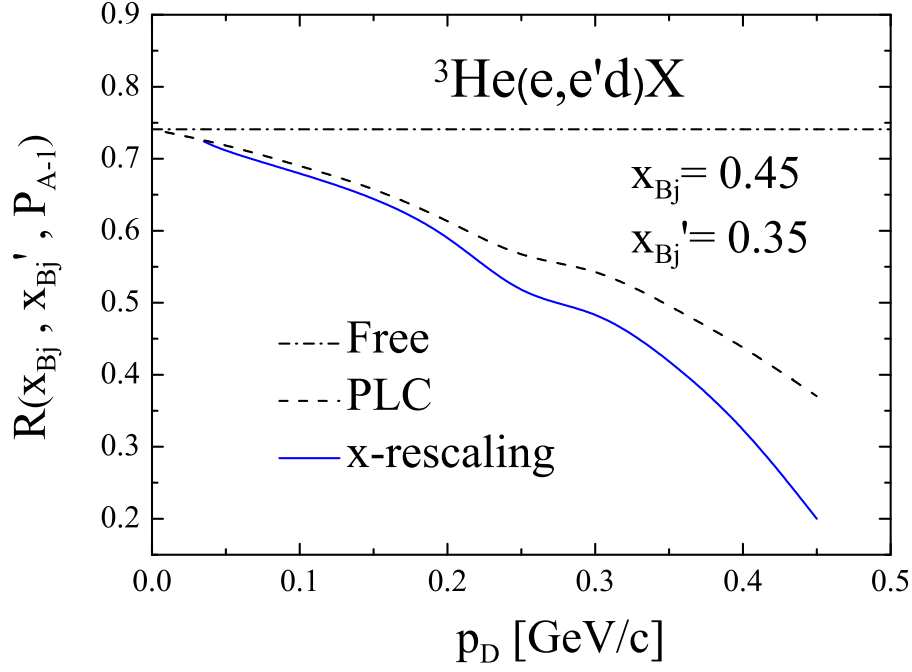}
\caption{The ratio given by Eq. (\ref{ratioa}) corresponding to  the process $^3He(e,e'd)X$  calculated at two values of the
Bjorken scaling variable $x_{Bj}$ and with different
nucleon structure functions. i)
 Free structure function (dot-dashed line): $F_2^{N/A} \left ( { x^A}  \right )
=F_2^{N/A} \left ( { x_{Bj}}  \right )$; ii) off mass-shell
(x-rescaling) structure function (full line): $F_2^{N/A} \left ( { x^{A}}\right ) =F_2^{N/A} \left
( { x_{Bj}/z_1^A}  \right )$ with $z_1^A= k_1 \cdot q/(m_N \,\nu)$ and $k_1^{0} = M_A -
\sqrt{{(M_{A-1}^{*})}^2+{\bf P}_{A-1}^2}$; iii)
 structure function with reduction of point-like configurations (PLC) in the medium depending upon
 the nucleon virtuality  $v({\bf k}_1,E)$ (Eq. (\ref{virtuality})) \cite{CKFS} (dashed line):
$F_2^{N/A} \left ( { x^{A}} \right )
=F_2^{N} \left ( { x_{Bj}/z_1^N}  \right )\cdot \delta_A(x_{Bj},v({\bf k}_1,E))$ with
$z_1^A= k_1 \cdot q/(m_N \,\nu)$ and   $k_1^{0} = M_A -
\sqrt{m_N^2+{\bf P}_{A-1}^2}$.}
\label{Fig6}
\end{figure}

 We thank Sebastian Kuhn for providing the experimental data that we used to produce Fig.~\ref{Fig3} and
 for many  useful discussions and suggestions. Discussions with Alberto Accardi, Kawtar  Afidi, Boris  Kopeliovich and Veronica Palli are
 also gratefully
 acknowledged. Calculations have been performed
at CASPUR facilities under the Standard HPC 2010 grant \textit
{SRCnuc}. One of us (L.P.K) thanks the Italian Ministry of Education,
 University and Research (MIUR), and  the Department of Physics, University
of Perugia, for support
through the program "Rientro dei Cervelli".

\newpage
\vskip 6cm

\newpage

\end{document}